# Intensity modulated proton arc therapy via geometry-based energy selection for ependymoma

Wenhua Cao, PhD[1], Yupeng Li, MSc[1], Xiaodong Zhang, PhD[1], Falk Poenisch, PhD[1], Pablo Yepes, PhD[1,2],

Narayan Sahoo, PhD[1], David Grosshans, MD, PhD[3], Susan McGovern, MD, PhD[3], G. Brandon Gunn, MD[3],

Steven J. Frank, MD[3], Xiaorong R. Zhu, PhD[1]

[1]Department of Radiation Physics, The University of Texas MD Anderson Cancer Center, Houston, Texas, USA

[2]Department of Physics and Astronomy, Rice University, Houston, Texas, USA

[3]Department of Radiation Oncology, The University of Texas MD Anderson Cancer Center, Houston, Texas, USA

**Corresponding Author:**

Yupeng Li, M.Sc.

Department of Radiation Physics, Unit 1150, The University of Texas MD Anderson Cancer Center

1515 Holcombe Blvd, Houston, Texas 77030

Tel: 713-794-4112

Email: yupeng_li@yahoo.com

Running Title: Proton arc therapy via energy selection

**Abstract**

**Purpose:** We developed and tested a novel method of creating intensity modulated proton arc therapy (IMPAT) plans that uses computing resources similar to those for regular intensity-modulated proton therapy (IMPT) plans and may offer a dosimetric benefit for patients with ependymoma or similar tumor geometries.

**Methods:** Our IMPAT planning method consists of a geometry-based energy selection step with major scanning spot contributions as inputs computed using ray-tracing and single-Gaussian approximation of lateral spot profiles. Based on the geometric relation of scanning spots and dose voxels, our energy selection module selects a minimum set of energy layers at each gantry angle such that each target voxel is covered by sufficient scanning spots as specified by the planner, with dose contributions above the specified threshold. Finally, IMPAT plans are generated by robustly optimizing scanning spots of the selected energy layers using a commercial proton treatment planning system. The IMPAT plan quality was assessed for four ependymoma patients. Reference three-field IMPT plans were created with similar planning objective functions and compared with the IMPAT plans.

**Results:** In all plans, the prescribed dose covered 95% of the clinical target volume (CTV) while maintaining similar maximum doses for the brainstem. While IMPAT and IMPT achieved comparable plan robustness, the IMPAT plans achieved better homogeneity and conformity than the IMPT plans. The IMPAT plans also exhibited higher relative biological effectiveness (RBE) enhancement than did the corresponding reference IMPT plans for the CTV in all four patients and brainstem in three of them.

**Conclusions:** The proposed method demonstrated potential as an efficient technique for IMPAT planning and may offer a dosimetric benefit for patients with ependymoma or tumors in close proximity to critical organs. IMPAT plans created using this method had elevated RBE enhancement associated with increased linear energy transfer in both targets and abutting critical organs.

## 1 | INTRODUCTION

Proton arc therapy with spot scanning has been shown to improve the treatment plan conformity and surrounding organ at risk (OAR) sparing[1-4]. However, it is limited by a high computational cost for treatment planning. Optimization of all scanning spots from a relatively large number of beam angles may require high memory usage and long computation time. Energy-switching time is another limiting factor; for a synchrotron-type proton accelerator in particular, energy switching may take up to seconds.[5-7] Sanchez-Parcerisa et al.[8] compared different methods of proton arc treatment planning that select one or two energies per beam angle based on water-equivalent range relative to the distal and proximal limits of the target region. However, this oversimplified energy selection strategy may not be sufficient for tumors with complex shape and heterogeneous anatomical surroundings in terms of clinically acceptable dose coverage. Ding et al[1] iteratively increased the sampling frequency of beam angles, during which process energy layers were redistributed to newly created angles. Then a filtration process was used to remove low-weighted energy layers. Robust optimization was integrated in the planning iterations. They reported on clinical applications of their proton arc therapy planning approach[9].

Brain tumors such as ependymoma often involve treatment targets located near or overlapping critical organs, such as the brainstem. Authors have reported that proton therapy may be associated with increased radiation necrosis[10,11]. Further analysis of differential damage induced by protons and photons using imaging changes as biomarkers suggested that increased linear energy transfer (LET) contributes to a higher incidence of radiation damage to brain tissue in the proton therapy cohort[12,13]. One of the strategies to mitigate the relative biological effectiveness (RBE) of high proton LET in treating brain tumors is to include an increasing number of beam angles in the treatment plan, with the hope of spreading out high-LET protons that are mostly located at the distal ends of the beams. This may lead one to intuitively expect proton arc therapy to naturally have this advantage to a great extent. Recent studies[14,15] have shown that proton arc therapy may offer greater flexibilities than conventional IMPT in limiting high-LET proton irradiation within the target by adopting an energy selection method that stops proton beams mostly at the interior target.

As part of the effort to develop the proton arc therapy capability at our center, we aimed to achieve a simplified workflow for proton arc therapy planning to encourage further works on this topic. Our hypothesis here is that proton arc therapy plans can be generated to potentially offer clinical benefits, with a simplified workflow, and reduced computing resource usage compared to the existing methodologies. Herein we report on our novel method of proton arc treatment planning based on an energy selection process, the only added step to the regular intensity-modulated proton therapy (IMPT) planning workflow, that we developed to create high-quality, robust proton arc therapy plans, potentially with limited computing resources. Here we refer to spot-scanning-based proton arc therapy as intensity modulated proton arc therapy or IMPAT. We also report on the *in silico* application of our method to treatment of ependymoma in four patients.

## 2 | METHODS AND MATERIALS

### 2.1 | Treatment Planning Optimizer

Our IMPAT planning method was developed and preliminarily implemented with in-house program modules, and the beam-line calculation and optimizer of the commercial proton treatment planning system (TPS) RayStation® (RaySearch Laboratories AB, Stockholm, Sweden). Our energy selection process relied on RayStation providing the initial full spectrum of energy layers that longitudinally cover the whole treatment target with proper spacings at each gantry angle. After energy selection, the RayStation optimizer produced reasonably acceptable solutions for our treatment planning tasks with a relatively large number of gantry angles but with a greatly reduced number of energy layers at each angle. Robust proton plan optimizations were applied with a setup uncertainty ±2 mm in x, y and z directions and range uncertainty ±2.5% for all clinical cases in this study. Robustness objectives were used for targets for robust coverage of targets.

**2.2 | Beam arrangement**

The proton arc was formed by consecutively placing gantry angles at 5° intervals to form a proper partial arc. For ependymoma treatment, start and end gantry angles of the partial arc were selected to avoid passing beams through optic organs, nasal and oral cavities, and cochleae. This is to achieve the best protection possible to those normal tissues and organs. To facilitate the assessment of IMPAT plan quality, the plans were compared side by side with reference three-field IMPT plans with the similar objective functions and optimization processes. IMPT beam angles were selected based on the similar principles to minimize passing beams through critical organs as well as to follow the target shape.

**2.3 | Geometry-based energy selection**

The water-equivalent path length was calculated for all dose voxels within the treatment target plus margins at each gantry angle by ray-tracing with computed tomography (CT) data. The resolution of the dose grid was set to 2 mm. The calibration data that map CT Hounsfield units to relative stopping power ratios to water were obtained from beam configurations of the TPS. For each dose voxel, our energy selection process collects all scanning spots with dose contributions above a specified fraction of the maximum spot dose at the Bragg peak (Figure 1). In the present study, a fractional threshold of 0.8 allowed for a sufficiently large pool of scanning spots with reasonable memory usage (<500 MB per patient for all four patients). The above data can be denoted as the dose deposition matrix D with its element Dij representing the relative dose received by dose voxel i from scanning spot j. Specifically, we call D the major dose deposition matrix to reflect the fact that only dose contributions above the threshold are included in D. Because only the major dose contributions from scanning spots had to be counted, dose contributions to dose voxels were estimated using lateral spot profiles modeled by the single Gaussian function according to the lateral distance between the voxel and the central axis of a scanning spot[16,17]. The sigma of the single Gaussian function was quantified based on in-air spot sizes, and in-medium multiple Coulomb scattering. The multiple Coulomb scattering was calculated using the generalized Highland approximation.[18,19] One may consider to take the proton plan robustness into consideration by including a set of beam perturbation scenarios during ray tracing and scanning-spot searching. However, this did not result in a noticeable improvement in our preliminary implementation and therefore was not applied in this study.

From the initial set of energy layers provided by the TPS , our energy selection module picks a subset of energy layers at each gantry angle such that the total number of scanning spots with dose contributions above a threshold fraction to each dose voxel meets or exceeds a specified minimum number, or minimum peak repetition. A minimum peak repetition of 7 was used in generating IMPAT plans for all four patients. With collections of scanning spots above the dose threshold, our energy selection is carried out in iterations. During each iteration, energy layers from all gantry angles are sorted according to the number of dose voxels that receive dose contributions above the fractional threshold from at least one scanning spot on the examined layer, and the energy layer with dose contributions above the threshold to the most dose voxels is selected. When sorting energy layers at each iteration, the energy layers selected in previous iterations are excluded from the sorting, as are the dose voxels that already satisfy the minimum peak repetition. The energy selection stops when all dose voxels satisfy the minimum peak repetition. Figure 2 illustrates the workflow of the above process.

## 2.4 | Patient study

Four ependymoma patients were selected to demonstrate IMPAT planning results using the proposed energy selection method. The prescription dose was 5400 cGy to be delivered in 30 fractions.

Raystation distinguishes between objective functions and constraints such that constraints must not be violated, whereas objective functions are to be minimized. In comparing the quality of the IMPAT and regular IMPT plans, the aim was to achieve optimal clinical target volume (CTV) coverage without violating the same maximum dose constraint on the brainstem. For this study, a constraint value of 5450 cGy was applied in plan optimization for all four patients. By maintaining the same brainstem constraint, the doses to other OARs were decreased to as much as possible without dramatically compromising the CTV coverage. When possible, CTV coverage was normalized such that the prescription dose covered 95% of the CTV while maintaining the similar brainstem dose.

Treatment plan quality was reported using dose-volume histograms (DVHs), plan robustness, the homogeneity index, and the conformity index. Plan robustness is shown qualitatively in band DVH graphs. The homogeneity index was calculated using the following formula described in Wu, et al.[20]:

$$HI = \frac{D_2 - D_{98}}{D_{prescription}} \times 100\%$$

where $D_{98}$ and $D_2$ are the minimum doses received by 98% and 2% of the target volume, respectively, and $D_{prescription}$ is the prescription dose. The conformity index or conformation number is calculated as described by Riet, et al.[21]:

$$CI = \frac{V_{T,p}}{V_T} \times \frac{V_{T,p}}{V_p}$$

where $V_{T,p}$ is the volume of the target receiving the prescription dose, $V_T$ is the target volume, and $V_p$ is the total volume receiving the prescription dose. Lower homogeneity index values represent more homogeneous target dose, whereas higher conformity index values represent better target dose conformity.

**2.5 | LET dependent RBE dose**:

LET distributions were calculated using a track-repeating Monte Carlo algorithm that was developed and validated by Yepes et al.[22,23] With a simplified formula, Unkelbach, et al.[24] modeled the RBE dose (b) approximately as follows:

$$b = (1 + cLET)d$$

or

$$b = d + cLETd$$

where $d$ is the physical dose and the LET dependency of the RBE factor is approximated as $1 + cLET$. With the constant $c$ set to 0.04 µm/keV, the RBE factor equals 1.1, a constant RBE factor adopted by many proton facilities, at the center of a spread-out Bragg peak of 5-cm modulation and 10-cm range where the dose-averaged LET is about 2.5 keV/µm. The part of the RBE dose enhanced by variable proton LET can be represented by term $cLETd$; the RBE dose enhancement was compared for the IMPAT and IMPT plans.

## 3 | RESULTS

The CTV volumes, CTV and brainstem overlaps, and field arrangements of IMPAT and IMPT plans for the patients in this study are listed in Table 1. The CTV and brainstem overlaps are listed in cc and percentages of the CTV volumes. The number of proton beams ranged from 33 to 37 for the four IMPAT plans, and the average number of energy layers ranged from 2.9 to 4.1 per gantry angle. The energy selection took 8.2, 7.4, 3.0, and 4.7 seconds for patient 1, 2, 3, and 4, respectively, on a desktop workstation with Intel® Core™ i7 CPU at 2.90 GHz.

Figure 3 shows a comparison of the transverse planar dose distributions for the IMPAT and reference IMPT plans. The prescription dose covered 95% of the CTV while similar maximum doses for the brainstem were maintained for all plans. Sparing of other OARs was mostly comparable for the IMPAT and IMPT plans (Figure 4). Figure 5 shows that the IMPAT and IMPT plans had comparable plan robustness according to the worst-case DVH curves and spread of band DVHs. IMPAT achieved homogeneity indices of 5.2%, 5.4%, 5.2% and 6.6% for patient 1, 2, 3, and 4, respectively. These plans were less heterogeneous than the reference IMPT plans, which had homogeneity indices of 7.8%, 8.6%, 7.0%, and 8.5%, respectively (Figure 6a). IMPAT plans had conformity indices of 0.665, 0.587, 0.573, and 0.501 for patient 1, 2, 3, and 4, respectively. These plans were more conformal than the reference IMPT plans, which had conformity indices of 0.628, 0.490, 0.454, and 0.443, respectively (Figure 6b).

Table 2 compares the RBE enhancement represented by term $cLETd$ for the CTV and brainstem in the IMPAT and reference IMPT plans. The IMPAT plans had greater RBE enhancement than did the reference IMPT plans in the CTV for all four patients, and in the brainstem for three of the four patients (Figure 7). The ratios of the mean RBE enhancements in the CTV for IMPAT to IMPT were 1.21, 1.11, 1.14, 1.15 in patients 1, 2, 3, 4, respectively. The ratios of D1% (the minimum value that the hottest 1% of the volume receives) of the brainstem RBE enhancement for IMPAT to IMPT were 0.92, 1.47, 1.34, 1.24 in patients 1, 2, 3, and 4, respectively.

## 4 | DISCUSSION

Proton treatment plan optimization is often a resource-intensive process in terms of computing memory and speed. IMPAT considerably magnifies this demand. Our novel IMPAT planning method first determines a minimum set of energy layers per gantry angle based on ray tracing, and single Gaussian lateral spot profiles, which is followed by proton treatment plan optimization. Therefore, it can be used to produce a clinically sound plan in a relatively small amount of time and with memory consumption no greater than that required for regular IMPT planning. A number of studies were focused on proton arc planning methods. Among them, a representative work by Ding el al.[1,9,25-27] included robust optimization that has become a required part of clinical planning procedures for many proton therapy centers. Our proposed method also included robust optimization as a necessary component in order to provide proton arc plans with clinically acceptable plan quality. Our work could be potentially valuable in providing an alternative planning workflow by reducing the total number of robust optimizations invoked between resampling and redistributions steps in Ding et al.'s method.

The initial full spectrum of energy layers at each gantry angle is crucial for the success of the energy selection. Although this was found not necessary for each proton arc planning task, the convergence of the energy selection iterations can be checked prior to the selection by counting for each target voxel how many scanning spots from the initial pool of energy layers are contributing spot doses above the specified dose threshold. The count of major contributing spots per target voxel should be greater (normally much greater) than the specified minimum peak repetition, which guarantees the energy selection terminates after finite iterations. If the count is small relative to the specified minimum peak repetition, the initial full spectrum of energy layers should be examined on whether they completely cover the entire treatment target, or spot spacings are too coarse.

The parameter choices of the minimum peak repetition and the fractional dose threshold were made by trial and error with a limited number of clinical cases tested. As more and more cases of various treatment sites are being tested, we expect these parameters can be optimized for specific treatment sites and scenarios. Generally speaking, as the minimum peak repetition is decreased, the decreased plan quality can be expected. As it is increased, however, the energy layers per gantry angle may increase rapidly and the delivery efficiency will be negatively impacted. The choice of the fractional dose threshold considered its impact on the energy selection efficiency, as well as memory usage. If the value is much smaller than 1.0 (e.g., 0.5 or 0.6), the energy selection suffers from too much noise from scanning spots contributing doses from the proximal parts of their single-spot dose profiles too far away from the Bragg peaks. If the value is too close to 1.0, a large number of energy layers may be selected for each field angle, or worse, the energy selection may fail to converge. Although the plan quality may improve to a certain degree with increased energy layers and scanning spots, it may start deteriorating due to the effect of "MU starvation"[28] or the optimizer becoming inefficient for the enormously large dimension of the problem. It is possible that at each iteration of the energy selection, multiple energy layers could make major spot dose contributions to the same count of target voxels. With our currently implementation, one layer is selected randomly among those equally contributing layers. Further improvement of our algorithm could be made by considering additional selection criteria, for example, to further improve OAR sparing.

The conformity that a proton treatment plan can achieve depends on multiple factors, such as the target volume and complexity of the target shape. For the patients in the present study, overlap of the CTV and brainstem may have limited the plan conformity. The conformity indices for both IMPAT and IMPT were highest for the patient with the least overlap. The difference in conformity between IMPAT and IMPT also was the smallest for that patient. Although it met clinical dose objectives generally adopted, sparing of cochleae was noticeably worse in the proton arc plan than in the reference IMPT plan for patient 3. Upon analyzing cochlea doses contributed from each field angle, the added field angles in the proton arc plan appeared to increase the chance of scanning spots being placed close to the cochleae for this patient. In terms of sparing an anatomic region, a proton optimizer generally does not work as effectively as not placing spots in or near that region at all.

In a regular IMPT plan, the optimizer often assigns heavy weights to energy layers that are of or close to the highest energy when each beam is examined individually. Those layers are commonly located at the edge of or even outside the targets. Hence, high-LET regions at the distal ends of heavy-weighted layers are often located outside the targets. Our energy selection process aims to cover targets with sufficient scanning spots with a minimum number of energy layers per beam angle. As a result, energies with ranges reaching around the middle of target depths often have a large number of scanning spots covering a relatively large cross-section of a tumor and tend to be selected with high priority. This is attributed to the fact that energy layers near distal and proximal depths normally have smaller numbers of scanning spots than do those at the middle depths for naturally shaped tumors. This likely contributed to the IMPAT plans having higher LET-associated RBE enhancement in the CTVs than did the corresponding reference IMPT plans. This effect was reported previously by Bertolet et al[14]. Due to overlap of the CTV and brainstem, the RBE enhancement was also elevated in the brainstem for all patients except patient 1, whose overlap was much smaller than that of the other three patients.

For the current proton therapy system equipped at our center (PROBEAT®, Hitachi, Ltd.), a proton beam is extracted at only one energy level within each cycle or spill. To switch to next energy, it takes 2.1 seconds to deaccelerate and accelerate protons again with a new spill. Hitachi is now supporting multi-energy extraction (MEE) in their recent models to improve delivery efficiency. In MEE, if there are protons left in the current spill after an energy level is delivered, the circulating protons remaining in the system can be deaccelerated to nearby energy levels and extracted again. One possible way of delivering the proton arc therapy may be to continuously deliver protons while gantry is rotating as in the volumetric modulated arc therapy in photon radiation therapy. As shown in Figure 8, for all four patients the proton energies selected by the proposed energy selection method congregated in narrow energy ranges for majority of field angles; proton energies also varied gradually between adjacent fields. These characteristics would naturally facilitate above mentioned possibilities of delivery techniques and may potentially benefit delivery efficiency of proton arc plans.

Scanning spot contributions were calculated in an in-house program module using ray tracing and lateral spot profile modeling with a single Gaussian function. The differences in ray tracing and spot profile modeling between this in-house system and the commercial TPS may have affected the optimality of our energy selection. The optimality and efficiency of the energy selection step could be further improved if the spot contribution calculation were an integrated part of the TPS.

We chose a gantry angle interval of 5° to sufficiently demonstrate the proposed planning method. No fundamental difficulty should prevent the method from working with finer or coarser gantry angle arrangements.

## 5 | CONCLUSIONS

The proposed IMPAT planning method is capable of creating plans of clinically acceptable quality for ependymoma patients. It has potential benefits of enhancing the conformity and homogeneity of target coverage, especially in close proximity to critical OARs. Furthermore, it allows for plan generation that is not much different than that using regular IMPT optimization processes by applying the proposed energy selection step beforehand with cost-efficient computing resource consumption. Further research is warranted to distribute the RBE enhancement, that is potentially associated with our IMPAT planning method, in targets and critical organs in a desirable manner.

**ACKNOWLEDGEMENTS**

The authors thank Donald Norwood of the Research Medical Library at the University of Texas MD Anderson Cancer Center for editing this manuscript.

**CONFLICT OF INTEREST**

None

**REFERENCES**

1. Ding X, Li X, Zhang JM, Kabolizadeh P, Stevens C, Yan D. Spot-Scanning Proton Arc (SPArc) Therapy: The First Robust and Delivery-Efficient Spot-Scanning Proton Arc Therapy. *International Journal of Radiation Oncology*Biology*Physics*. 2016/12/01/ 2016;96(5):1107-1116. doi:https://doi.org/10.1016/j.ijrobp.2016.08.049

2. Rah J-E, Kim G-Y, Oh DH, et al. A treatment planning study of proton arc therapy for para-aortic lymph node tumors: dosimetric evaluation of conventional proton therapy, proton arc therapy, and intensity modulated radiotherapy. *Radiation Oncology*. 2016/10/21 2016;11(1):140. doi:10.1186/s13014-016-0717-4

3. Seco J, Gu G, Marcelos T, Kooy H, Willers H. Proton Arc Reduces Range Uncertainty Effects and Improves Conformality Compared With Photon Volumetric Modulated Arc Therapy in Stereotactic Body Radiation Therapy for Non-Small Cell Lung Cancer. *International Journal of Radiation Oncology*Biology*Physics*. 2013/09/01/ 2013;87(1):188-194. doi:https://doi.org/10.1016/j.ijrobp.2013.04.048

4. Hecht A. SU-FF-T-136: Improved Dose Characteristics From Proton Arc Therapy. https://doi.org/10.1118/1.3181610. *Med Phys*. 2009/06/01 2009;36(6Part10):2551-2551. doi:https://doi.org/10.1118/1.3181610

5. Li Y, Kardar L, Li X, et al. On the interplay effects with proton scanning beams in stage III lung cancer. *Med Phys*. 2014;41(2):021721-1. doi:doi:http://dx.doi.org/10.1118/1.4862076

6. Klein H-U, Baumgarten C, Geisler A, et al. New superconducting cyclotron driven scanning proton therapy systems. *Nuclear Instruments and Methods in Physics Research Section B: Beam Interactions with Materials and Atoms*. 2005/12/01/ 2005;241(1):721-726. doi:https://doi.org/10.1016/j.nimb.2005.07.123

7. Cao W, Lim G, Liao L, et al. Proton energy optimization and reduction for intensity-modulated proton therapy. *Phys Med Biol*. 2014;In press

8. Sanchez-Parcerisa D, Kirk M, Fager M, et al. Range optimization for mono- and bi-energetic proton modulated arc therapy with pencil beam scanning. *Phys Med Biol*. 2016/10/14 2016;61(21):N565-N574. doi:10.1088/0031-9155/61/21/n565

9. Li X, Liu G, Janssens G, et al. The first prototype of spot-scanning proton arc treatment delivery. *Radiother Oncol*. 2019/08/01/ 2019;137:130-136. doi:https://doi.org/10.1016/j.radonc.2019.04.032

10. Indelicato DJ, Flampouri S, Rotondo RL, et al. Incidence and dosimetric parameters of pediatric brainstem toxicity following proton therapy. *Acta Oncologica*. 2014/10/01 2014;53(10):1298-1304. doi:10.3109/0284186X.2014.957414

11. Weber DC, Malyapa R, Albertini F, et al. Long term outcomes of patients with skull-base low-grade chondrosarcoma and chordoma patients treated with pencil beam scanning proton therapy. *Radiother Oncol*. 2016/07/01/ 2016;120(1):169-174. doi:https://doi.org/10.1016/j.radonc.2016.05.011

12. Mohan R, Peeler CR, Guan F, Bronk L, Cao W, Grosshans DR. Radiobiological issues in proton therapy. *Acta Oncologica*. 2017/11/02 2017;56(11):1367-1373. doi:10.1080/0284186X.2017.1348621

13. Peeler CR, Mirkovic D, Titt U, et al. Clinical evidence of variable proton biological effectiveness in pediatric patients treated for ependymoma. *Radiother Oncol*. 2016/12/01/ 2016;121(3):395-401. doi:https://doi.org/10.1016/j.radonc.2016.11.001

14. Bertolet A, Carabe A. Proton monoenergetic arc therapy (PMAT) to enhance LETd within the target. *Physics in Medicine & Biology*. 2020/08/19 2020;65(16):165006. doi:10.1088/1361-6560/ab9455

15. Carabe A, Karagounis IV, Huynh K, et al. Radiobiological effectiveness difference of proton arc beams versus conventional proton and photon beams. *Physics in Medicine & Biology*. 2020/08/31 2020;65(16):165002. doi:10.1088/1361-6560/ab9370

16. Li Y, Zhang X, Mohan R. An efficient dose calculation strategy for intensity modulated proton therapy. *Phys Med Biol*. Feb 2011;56(4):N71-N84. doi:10.1088/0031-9155/56/4/n03

17. Li Y, Zhu RX, Sahoo N, Anand A, Zhang XD. Beyond Gaussians: a study of single-spot modeling for scanning proton dose calculation. Article. *Phys Med Biol*. Feb 2012;57(4):983-997. doi:10.1088/0031-9155/57/4/983

18. Gottschalk B. On the scattering power of radiotherapy protons. Article. *Med Phys*. Jan 2010;37(1):352-367. doi:10.1118/1.3264177

19. Hong L, Goitein M, Bucciolini M, et al. A pencil beam algorithm for proton dose calculations. Article. *Phys Med Biol*. Aug 1996;41(8):1305-1330.

20. Wu Q, Mohan R, Morris M, Lauve A, Schmidt-Ullrich R. Simultaneous integrated boost intensity-modulated radiotherapy for locally advanced head-and-neck squamous cell carcinomas. I: dosimetric results. *International Journal of Radiation Oncology*Biology*Physics*. 2003/06/01/ 2003;56(2):573-585. doi:https://doi.org/10.1016/S0360-3016(02)04617-5

21. Riet Avt, Mak ACA, Moerland MA, Elders LH, van der Zee W. A conformation number to quantify the degree of conformality in brachytherapy and external beam irradiation: Application to the prostate. *International Journal of Radiation Oncology*Biology*Physics*. 1997/02/01/ 1997;37(3):731-736. doi:https://doi.org/10.1016/S0360-3016(96)00601-3

22. Yepes PP, Guan F, Kerr M, et al. Validation of a track-repeating algorithm versus measurements in water for proton scanning beams. *Biomedical Physics & Engineering Express*. 2016/05/27 2016;2(3):037002. doi:10.1088/2057-1976/2/3/037002

23. Yepes PP, Mirkovic D, Taddei PJ. A GPU implementation of a track-repeating algorithm for proton radiotherapy dose calculations. *Phys Med Biol*. 2010/11/12 2010;55(23):7107-7120. doi:10.1088/0031-9155/55/23/s11

24. Unkelbach J, Botas P, Giantsoudi D, Gorissen BL, Paganetti H. Reoptimization of Intensity Modulated Proton Therapy Plans Based on Linear Energy Transfer. *International Journal of Radiation Oncology*Biology*Physics*. 2016/12/01/ 2016;96(5):1097-1106. doi:https://doi.org/10.1016/j.ijrobp.2016.08.038

25. Ding X, Li X, Qin A, et al. Have we reached proton beam therapy dosimetric limitations? – A novel robust, delivery-efficient and continuous spot-scanning proton arc (SPArc) therapy is to improve the dosimetric outcome in treating prostate cancer. *Acta Oncologica*. 2018/03/04 2018;57(3):435-437. doi:10.1080/0284186X.2017.1358463

26. Li X, Kabolizadeh P, Yan D, et al. Improve dosimetric outcome in stage III non-small-cell lung cancer treatment using spot-scanning proton arc (SPArc) therapy. *Radiation Oncology*. 2018;13(1):1-9.

27. Liu G, Li X, Qin A, et al. Improve the dosimetric outcome in bilateral head and neck cancer (HNC) treatment using spot-scanning proton arc (SPArc) therapy: a feasibility study. *Radiation Oncology*. 2020/01/30 2020;15(1):21. doi:10.1186/s13014-020-1476-9

28. Zhu XR, Sahoo N, Zhang X, et al. Intensity modulated proton therapy treatment planning using single-field optimization: The impact of monitor unit constraints on plan quality. *Med Phys*. Mar 2010;37(3):1210-1219. doi:10.1118/1.3314073

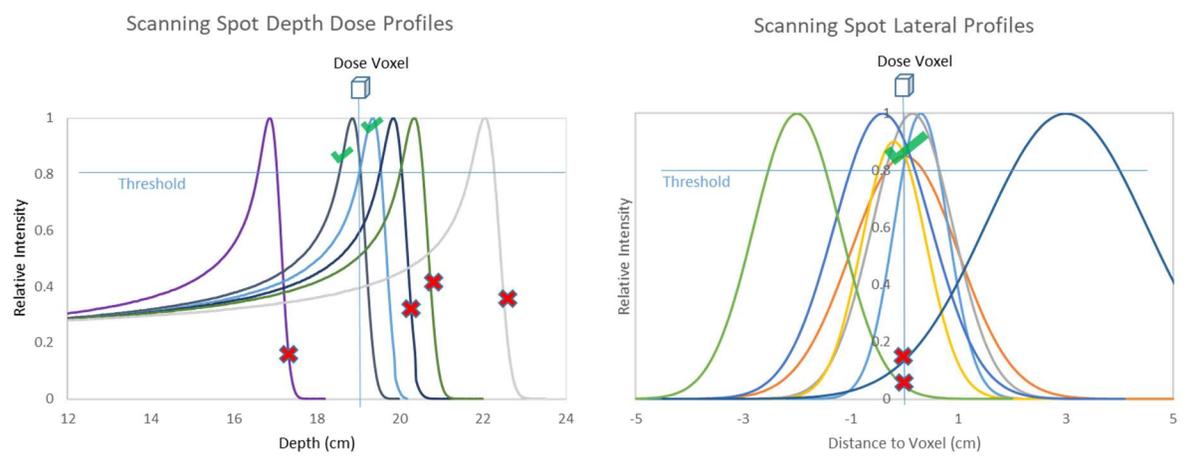

**FIGURE 1** For each target dose voxel, only scanning spots with dose contributions above a specified fractional threshold of the maximum spot dose (at the Bragg Peak) were counted in the energy selection process. Illustrated in the views of depth dose and lateral spot profiles respectively, green checkmarks indicate spots that are counted in the energy selection, and red cross marks indicate spots that are excluded in the process.

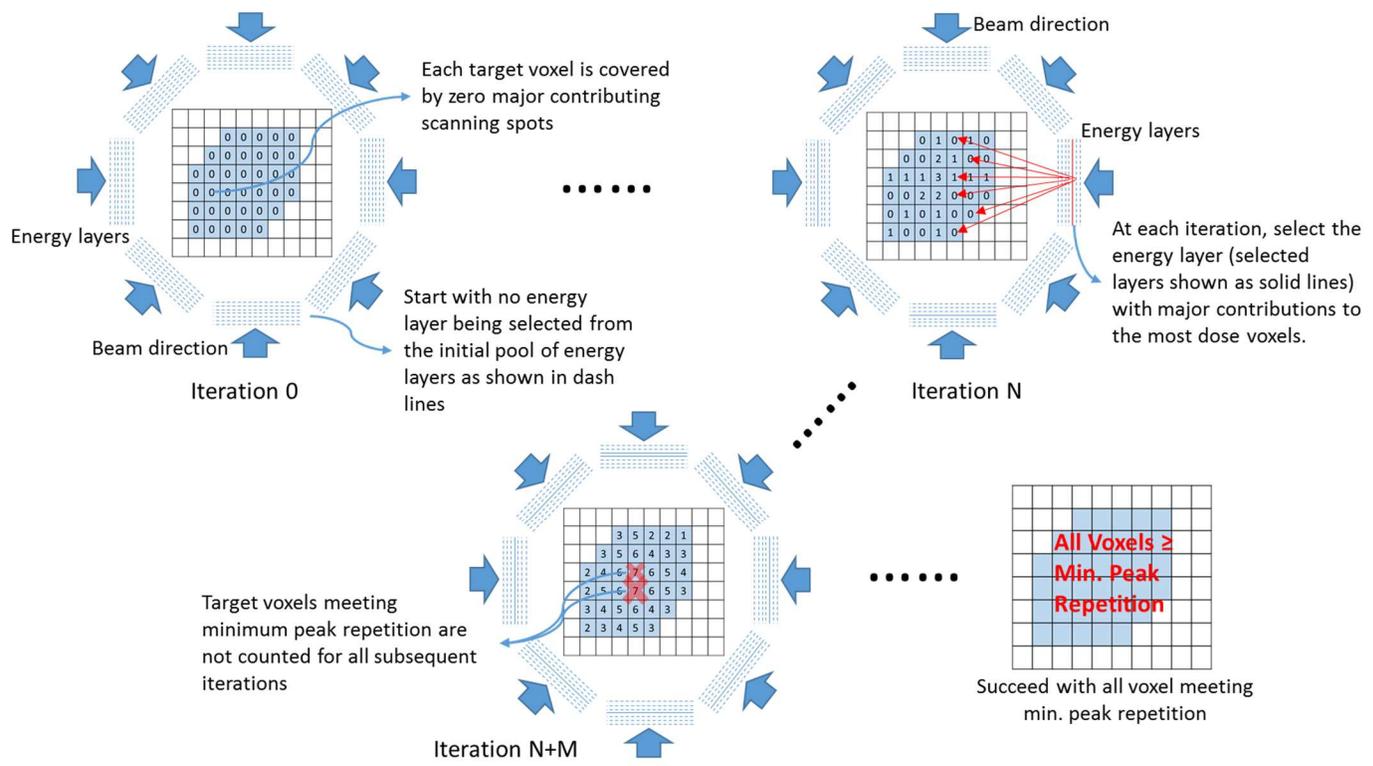

**FIGURE 2** Geometry-based energy selection workflow.

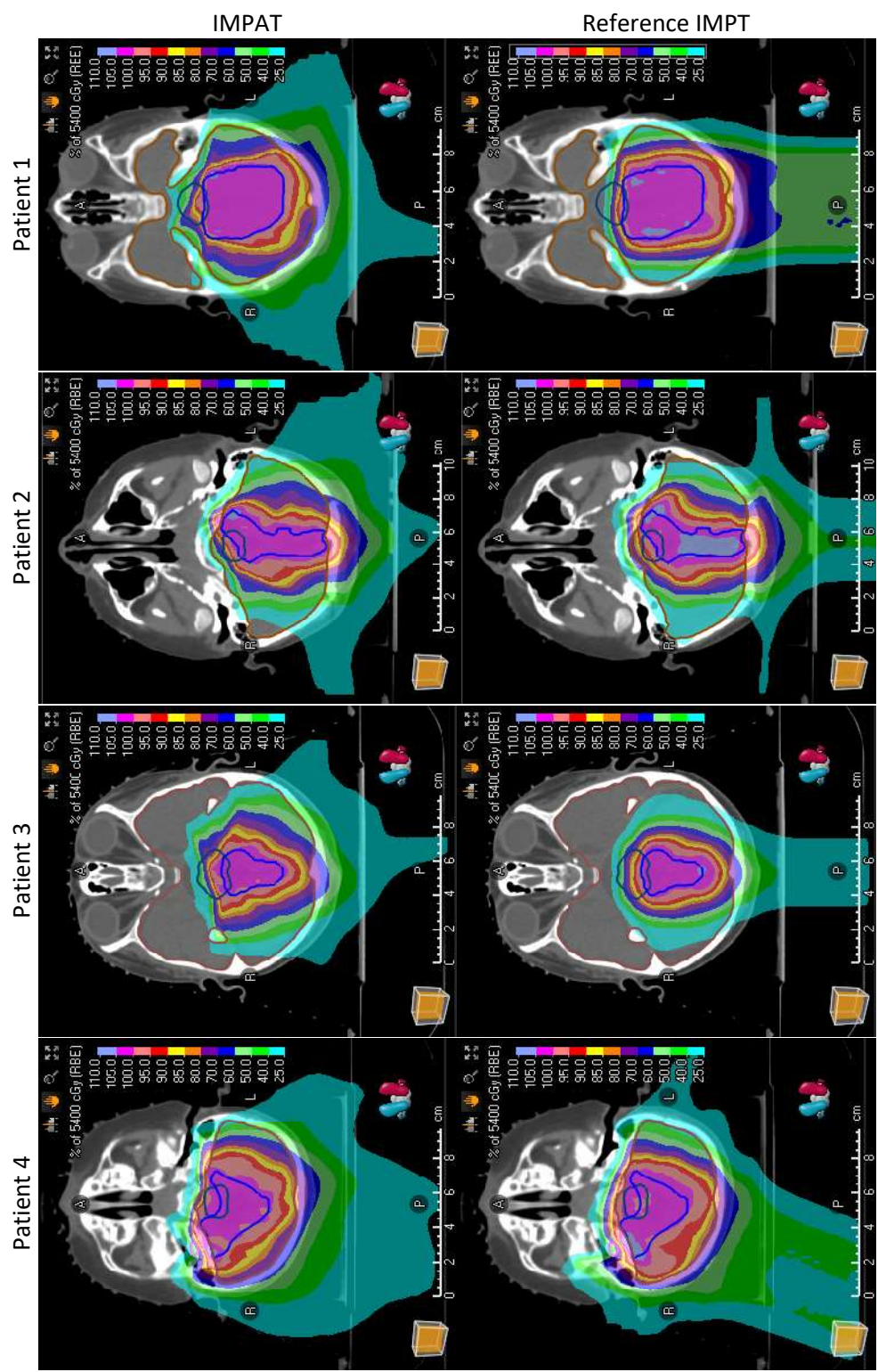

**FIGURE 3** Transverse planar dose distributions at the isocenters in the four study patients.

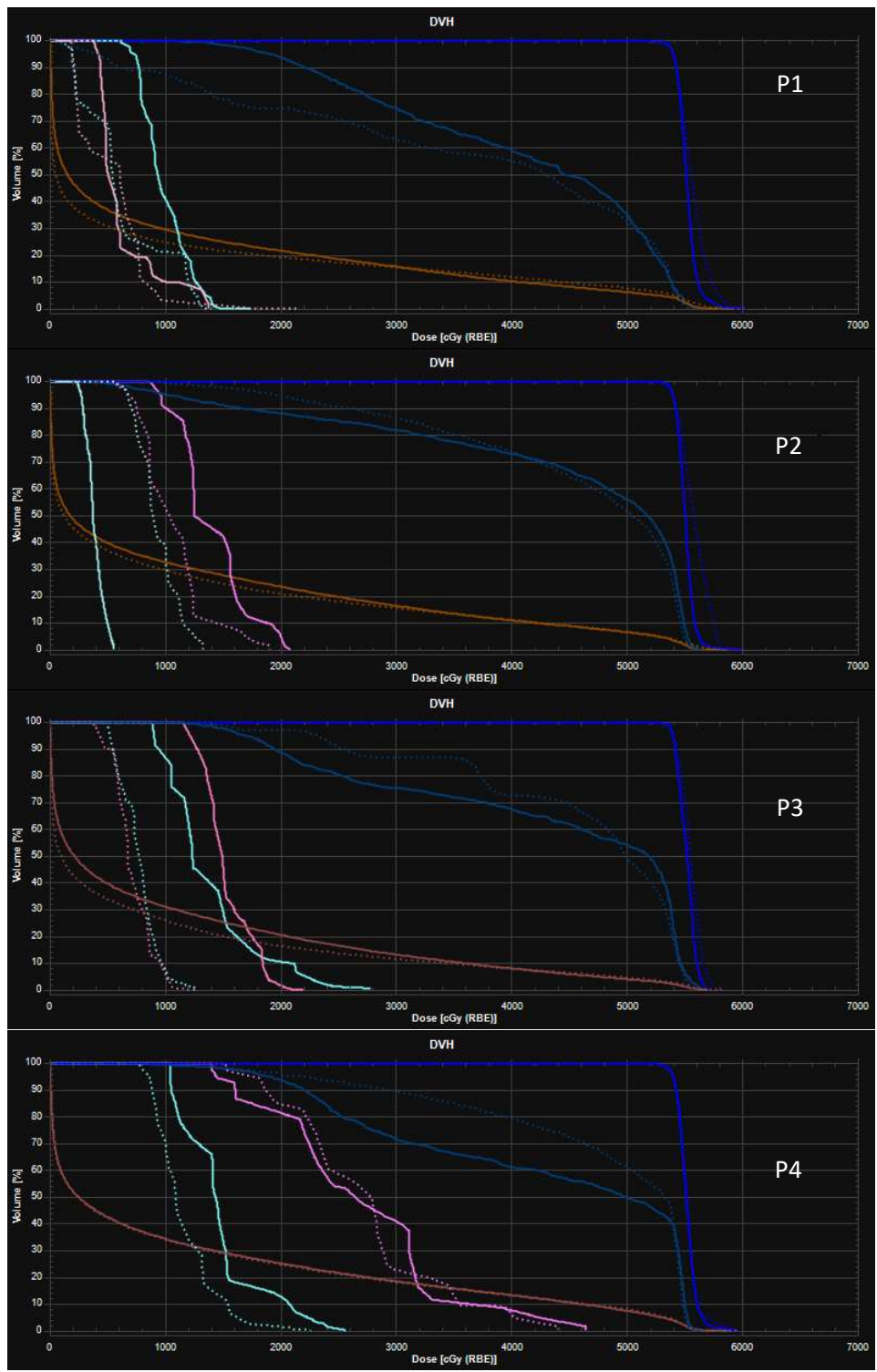

**FIGURE 4** Comparison of nominal DVHs for the IMPAT (solid lines) and IMPT (dotted lines) plans in the four study patients (P1-P4), including the DVHs for the CTV (cobalt), brainstem (blue), brain (brown), left cochlea (pink), and right cochlea (cyan).

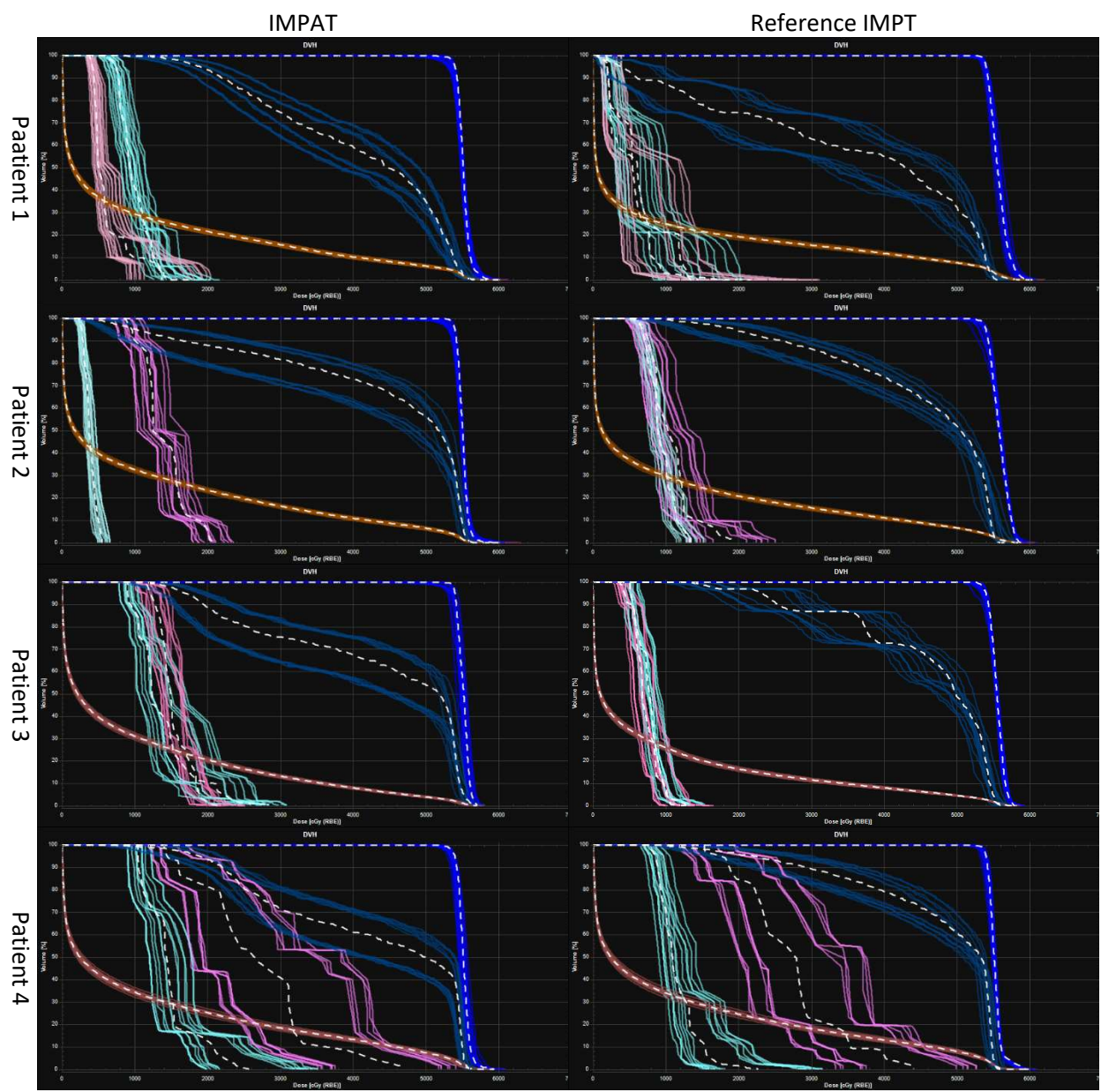

**FIGURE 5** Band DVHs for perturbed dose distributions under 12 scenarios (shifts of ±2 mm on the x, y, and z axes and ±2.5% on the range) in the four study patients. The structures shown are the CTV (cobalt), brainstem (blue), brain (brown), left cochlea (pink), and right cochlea (cyan). The white dashed lines represent nominal DVHs for corresponding structures.

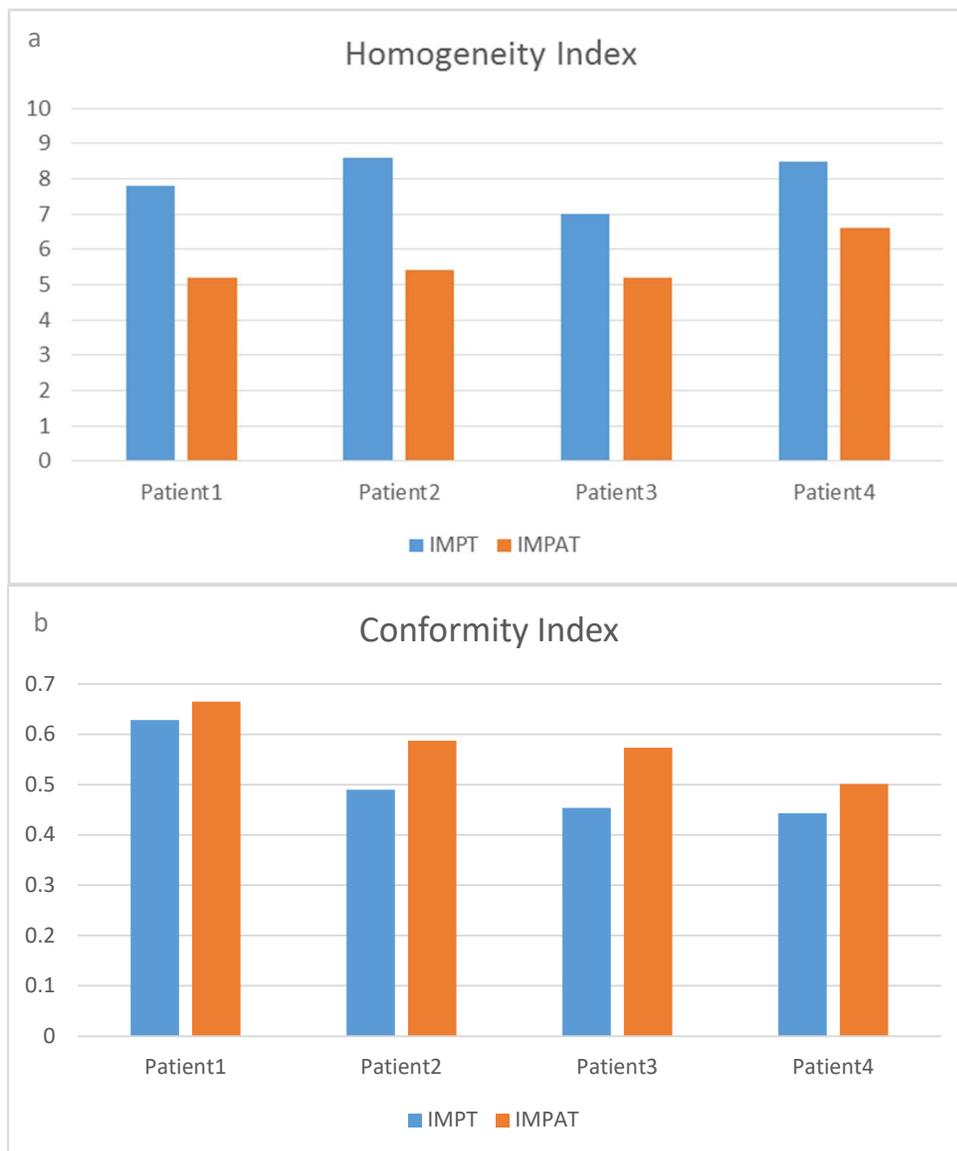

**FIGURE 6** Homogeneity indices (a) and conformity indices (b) for IMPAT (orange or right in the pairs) and IMPT (blue or left) plans. Lower homogeneity index values represent more homogeneous target dose. Higher conformity index values represent more conformal dose.

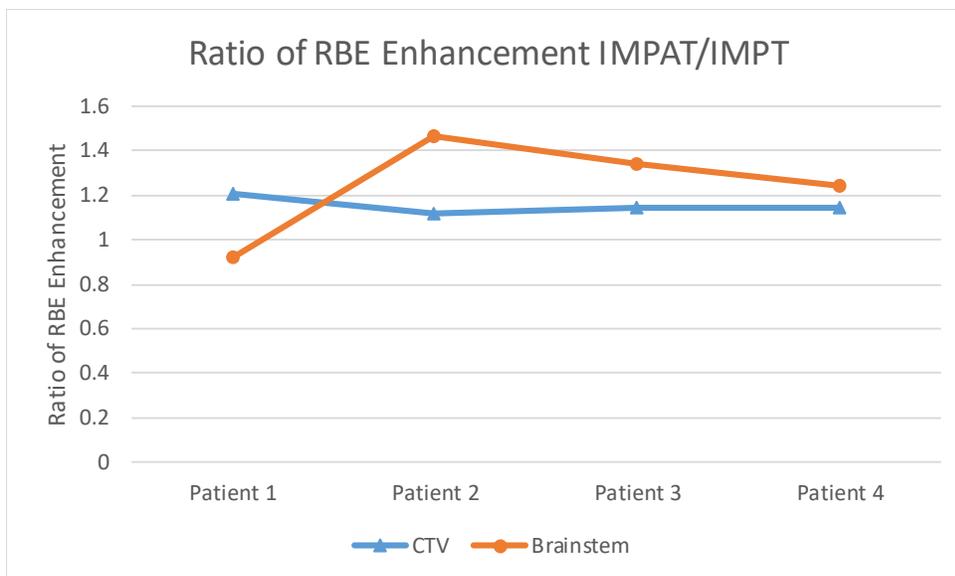

**FIGURE 7** Ratio of RBE enhancements for the IMPAT to reference IMPT plans in the CTV (ratio of mean enhancements in blue triangles) and brainstem (ratio of D1% in orange circles) in the four study patients.

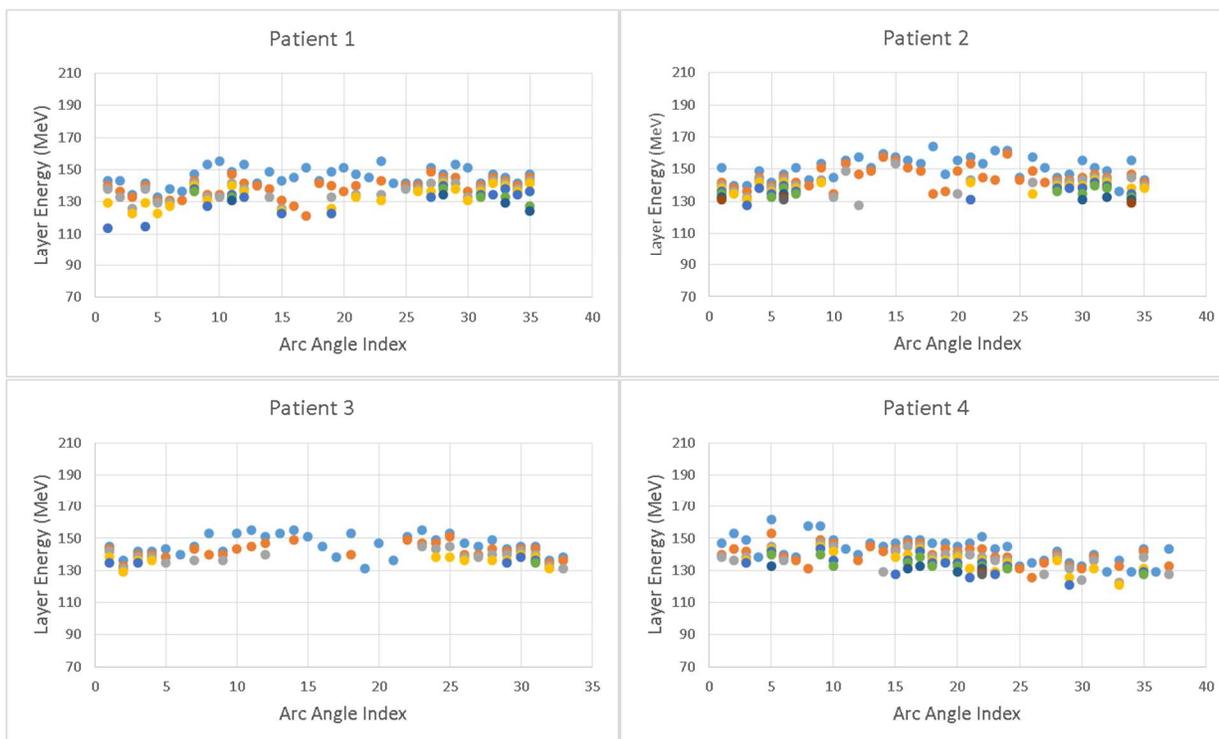

*FIGURE 8 The proton energies (as shown by dots in various colors) picked by the geometry-based energy selection process congregated in narrow energy ranges for majority of field angles; proton energies also varied relatively continuously between adjacent fields. These characteristics may potentially improve delivery efficiency of proton arc plans.*

**TABLE 1** CTV volumes, CTVs and brainstems overlaps in cc and in percentages of the CTV volumes, IMPAT arc angle ranges in a clockwise (CW) direction, total numbers of gantry angles in IMPAT plans, average numbers of layers per gantry angle of IMPAT plans, and IMPT gantry angles.

|  | Patient 1 | Patient 2 | Patient 3 | Patient 4 |
|---|---|---|---|---|
| CTV volume (cc) | 42.96 | 43.66 | 17.58 | 25.67 |
| CTV/brainstem overlap (cc/%) | 0.42/0.9 | 5.46/12.6 | 2.20/12.5 | 3.14/12.2 |
| IMPAT arc range CW (degrees) | 95-265 | 95-265 | 100-260 | 110-290 |
| Number of IMPAT angles | 35 | 35 | 33 | 37 |
| Average number of layers per IMPAT angle | 4.1 | 4.0 | 2.9 | 4.0 |
| IMPT angles (degrees) | 100, 180, 260 | 100, 180, 260 | 100, 180, 260 | 110, 200, 290 |

**TABLE 2** Comparison of RBE enhancement in cobalt cGy equivalent (CcGE) represented by term cLETd in the CTV and brainstem in the IMPAT and regular IMPT plans.

| | CcGE | | | |
|---|---|---|---|---|
| RBE enhancement | Patient 1 | Patient 2 | Patient 3 | Patient 4 |
| CTV mean | | | | |
|   IMPAT | 868.7 | 816.4 | 900.0 | 843.0 |
|   IMPT | 718.3 | 733.3 | 787.4 | 735.1 |
| Brainstem D1% | | | | |
|   IMPAT | 1429.4 | 1680.9 | 1691.0 | 1483.5 |
|   IMPT | 1548.1 | 1144.3 | 1261.7 | 1193.0 |